\documentclass[10pt, a4paper, notitlepage]{article}

\usepackage{tgpagella}  
\usepackage{mathpazo}   

\usepackage{geometry}
\usepackage{amsmath}
\usepackage{graphicx}
\usepackage[numbers]{natbib}
\bibpunct{(}{)}{;}{a}{,}{,}
\usepackage[dvipsnames]{xcolor}

\usepackage[colorlinks=true,
  linkcolor=CadetBlue,
  citecolor=CadetBlue,
  urlcolor=CadetBlue]{hyperref}
\usepackage{url}

\usepackage{multirow}
\usepackage{tabularx}
\usepackage{booktabs}

\usepackage{sectsty}
\sectionfont{\fontsize{12}{12}\selectfont}
       \subsectionfont{\fontsize{10}{10}\selectfont}

\begin{document}

\title{Do Cyber Capabilities and Cyber Power Incentivize International Cooperation?}
\author{\normalfont Jukka Ruohonen \vspace{4pt}\\ University of Turku \\ \texttt{juanruo@utu.fi}}

\maketitle

\begin{abstract}
This paper explores a research question about whether defensive and
offensive cyber security power and the capabilities to exercise the power
influence the incentives of nation-states to participate in bilateral and
multilateral cooperation (BMC) through formal and informal agreements,
alliances, and norms. Drawing from international relations in general and
structural realism in particular, three hypotheses are presented for assessing
the research question empirically: (i)~increasing cyber capability lessens the
incentives for BMC; (ii) actively demonstrating and exerting cyber power
decreases the willingness for BMC; and (iii)~small states prefer BMC for cyber
security and politics thereto. According to a cross-country dataset of $29$
countries, all three hypotheses are rejected. Although presenting a ``negative
result'' with respect to the research question, the accompanying discussion
contributes to the state-centric cyber security research in international
relations and political science.
\vspace{25pt}
\end{abstract}

\section*{Introduction}

This paper explores a research question about whether ``cyber capabilities'' and
``cyber power'', in particular, incentivize nation-states to participate in
bilateral and multilateral cooperation. The exploration is carried out with a
``cross-country'' empirical analysis operating at a ``macro-political''
level. In other words, the unit of analysis is a whole country. This
clarification opens a Pandora's box of well-known problems that should be
acknowledged before proceeding any further. The macro-political level entails
comparisons across ``the most complicated set of social units we know, total
societies; and generalizations do not come easily''
\citep[p.~113]{Verba67}. There are only a limited number of countries in the
world. Thus, the ``population'' is limited in statistical terms, but, in more
qualitative terms, even a comparison across two, three, or a few countries is
often challenging. So, to clear the muddy waters: the paper does not attempt to
generalize. Nor are causal claims made. The research approach is exploratory,
and the empirical dataset examined is based on different ``indices'' on ``cyber
security''.

Macro-political indices are notoriously difficult to construct. Why? To
understand the question, one only has to think about the two words: macro and
political. One thought arising immediately is that quantifying a macro-level
index for a whole country is often a political act. When politics are involved,
uncertainty increases. Mistakes happen, and mistakes are more prone to happen
when operating with concepts that cannot be defined and quantified
rigorously. Democracy is among these concepts, rule of law is among these
concepts, and cyber power is among these concepts. All these concepts are
difficult to define and quantify. Depending on a context,
indicators can be viewed as facts, proxies, predictors, diagnoses, targets for
reform, or conceptual frameworks, and depending on a given viewpoint, these are
affected by stakeholder incentives, comparability issues across countries,
reporting of uncertainty, and other problems~\citep{Botero11}. This alloy makes
it easy to mount attacks against comparative cross-country research. However,
much of the conventional criticism is almost universal criticism with which many
empirical sciences can be attacked~\citep{Swank07}. Despite the problems and the
criticism, macro-political indices can be constructed. Mistakes are
corrected. Macro-political indices are also needed for policy-making. In fact,
there are many international organizations, such as the Organisation for
Economic Co-operation and Development (OECD), devoted to cross-country data.

Old concepts are backed by mature measurement methodologies. Although these
often amount to different checklists, the lists are still encompassing and
increasingly sharp and well-assembled. But cyber power is not a mature
concept. Nor is cyber security. It is therefore understandable that there were
no macro-political indices for these concepts until recently. This lack has
contributed to a lack of comparative cross-country research. Although there are
some existing empirical studies~\citep{Holt16, Makridis19, Mezzour18}, these are
outliers even when compared to closely related research domains, such as
cross-country analysis of telecommunications, Internet adoption, and similar
topics. Given recent improvements with indices, however, the situation is likely
to change. This paper contributes to the nurturing of such a potential change.

Though, there is still even a debate on what cyber security actually is. To
provide a high-level overview of the terminology involved, cyber security is
often framed with the domain on which it operates (``cyber space''), offensive
and protective actions in this space, and the consequences from these
actions~\citep{Rackeviciene20}. This high-level terminological overview serves
its purpose: the cross-country focus is on the offensive capabilities and
actions on one hand, and the protective measures adopted by nation-states on the
other. The consequences are asserted to manifest themselves as increased desire
for international cooperation---or lack thereof. This focus places the paper
firmly into a so-called state-centric domain of cyber security
research~\citep[cf.][]{Eriksson09b, Ruohonen20EJSR}. The comparative
macro-political approach frames the paper's scope further; the word politics
underlines the presence of cyber security policies. Among other things, such as
economic realities, these policies are affected by technology, science, and
international (power) politics as well as domestic
politics~\citep{Cavelty19a}. This complex constellation defines the concept of
``cyber security politics''. Domestically, there are polities within which such
politics are made, but international cyber security politics have long been in
dire straits. Despite many initiatives, both laws and norms are immature at best
in the (international) cyber space. Although some of the initiatives have
matured into international treaties, such as the Budapest Convention, many of
the initiatives have been a ``mess in New York''~\citep[p.~298]{Maurer20},
composed of ``endless talkshops''~\citep{Register20a} rather than genuine
international treaty negotiations. This international chaos provides the
motivation to examine the question whether offensive and protective cyber
actions incentivize cooperation. The concept of ``cyber power'' can be defined
via these actions as the exercise, maintenance, or demonstration of power. The
``cyber capability'' for these actions originates from the combination of
offensive and protective potential, economy, technology, science, and the
domestic and international cyber security politics. With these clarifications, the hypotheses for the empirical exploration can be stated.

\section*{Hypotheses}

The preceding discussion about various ``cyber concepts'' aligns with classical
notions and theories in international relations. When keeping in \textit{mind}
the state-centric straitjacket, \citeauthor{Waltz79}'s
\citeyearpar[pp.~88--89]{Waltz79} famous theory seems pertinent as a starting
point: domestic politics occur in a centralized and hierarchical polity, whereas
international politics are organized anarchically without a common
polity. International relations do not differ from the cyber space in this
regard. There is no ``world government'' in either domain. Also other elements
in the Waltz's theory seem applicable as a heuristic driver toward developing
hypotheses. In particular, his notion of ``structure'' in international
relations---despite the anarchy (here, for helpful reading, see
\citealt{Lechner17}), provides a decent analytical vehicle. It has three layers:
\textit{ordering principles} (e.g., to move from an anarchic to a hierarchical
order changes a system's structure), \textit{functional differentiation} between
units, and the \textit{distribution of capabilities}
\citep[pp.~100--101]{Waltz79}. Changes in the distribution of capabilities may
change both anarchic and hierarchical systems; however, changes in the
functional differentiation cannot change an anarchic system because there is no
differentiation to begin with. How these structural layers can be used to
elaborate recent developments in the cyber space and cyber security politics?

To begin the elaboration, it must be assumed that a change in the ordering
principles is occurring; otherwise the research question would not make
sense. Having locked this assumption, it makes sense to start from the
functional differentiation in the cyber space, and particularly in the Internet
and its governance. For a long time, the governance of the Internet was highly
functionally differentiated. And it was never an anarchy. The governance of the
Internet mainly revolved around different loosely organized, non-state networks,
standardization organizations, committees, non-governmental organizations, and
the like, composed of engineers, scientists, hobbyists, and the like. These
governance bodies were---and still are---characteristically non-state, even
though they often received funding from governments and companies, and were
under their loose oversight. In fact, the whole concept of governance is often
understood to mean governing without the necessity of a
government~\citep{Rhodes96, Wachaus14}. In the Internet governance literature
this type of governing became known as a multi-stakeholder
governance~\citep{Strickling17}. However, cracks started to show in this type of
governance model as the 2010s progressed. While it was still easy to reject
state-centric explanations about a decade ago~\citep{Mueller08}, in the late
2010s the arguments against state-centric governance had turned defensive and
even pessimistic~\citep{Mueller19}. The ``militarization'' of the cyber
space~\citep{Deibert13} was one but not the only reason. Also the ``structure''
of the Internet centralized to the hands of a few large multinational companies.

Again, how to interpret these changes against Waltz's basic theoretical
prem\-ises?  At the risk of an overstatement, it could be said that states
and capitalism brought the anarchy \textit{of} international relations to the
cyber space. And with anarchy came the conventional speculations. Although the
Internet has so far shown remarkable resilience, discussions about hegemonic
power started to appear in the academic literature~\citep{RovnerMoore17}. In
2020 pundits and commentators speculated openly about the apparent
technology-related international power politics and the potential splintering of
the Internet and its governance into different regional
camps~\citep{CircleID20a}. At the same time, offensive cyber actions continued
unabated in the international anarchy. These developments awaken the Waltz's
theory. Thus, to continue with the theoretical premises: states at minimum seek
their self-preservation and at maximum seek a hegemonic position through their
domestic polities (i.e., in the present context, by increasing their cyber
capabilities) and international endeavors (e.g., by creating alliances or
weakening the alliances of others), such that a balance of power is achieved
only if all states recognize the same rules and play with the same minimum
bets~\citep[pp.~118--120]{Waltz79}. While acknowledging the limitations of this
simplification \citep{Blachford20}, these basic premises of structural realism
allow to posit the first~hypothesis:
\begin{description}
\item[\normalfont{H}$_1$]{\textit{Increasing cyber capability decreases
    willingness for BMC.}}
\end{description}

Note that the hypothesis is postulated with a negative correlation (i.e., due to
the hegemony assumption), although a reverse statement might be equally
plausible; increasing cyber capabilities of some states may incentivize other
states to form alliances with them, for instance. In contrast to this slight
ambiguity, a more direct hypothesis is available by assuming that active
demonstration \textit{and} use of cyber power signal a lack of strong incentives
for international cooperation:
\begin{description}
\item[\normalfont{H}$_2$]{\textit{Demonstrating and exerting cyber power decreases willingness for BMC.}}
\end{description}

Both $\textmd{H}_1$ and $\textmd{H}_2$ are exploratory yet still
theory-motivated hypotheses. To move toward the actual context, the 2010s was
also a period during which different ``cyber norms'' were actively pursued. For
the present purposes, such norms can be defined as ``non-binding conventions or
a standard of appropriate behavior about how a class of actors should act'',
such that, ``over time and when they provide order, stability, and security,
they are often codified into law'' \citep[p.~335]{Ryan18}. Partially due to the
anarchy, whether perceived or real, much of the intellectual fervor behind this
``norm-building'' drew from international relations and military
logic~\citep{Mueller19}. Even today---for some establishments, stakeholders, and
pundits, the ``hope is that the emergence of a Westphalian cyber-order will
bring back the certainty of the Cold War'' \citep[p.~196]{Balzacq16}. To this
end, indeed, it was common in the 2010s to formulate different general
principles for appropriate state behavior in the cyber space. For instance,
territorial sovereignty should be honored; every state has a right to build
cyber capabilities; criminal law should apply; all states have a right to
self-defense; and so forth~\citep{Tikk11}. Many analogous general principles
were codified into the so-called Tallinn Manual, released in 2013 and later
updated in 2017~(for background see \citealt{vonHeinegg14}). Alongside such
informal guidebooks came also more general theories such as deterrence; the idea
that escalating cyber conflicts between states could be mitigated with
international norms that, at minimum, make cyber attacks politically
costly~\citep{Taddeo17b}. Although it remains unclear whether deterrence tactics
work in the cyber space~\citep{Register20b, Nye17}, many states have recently
adopted these into their cyber security strategies~\citep{Olejnik20}, some of
which may also include an aggressive ``strike-back'' deterrence variant.

To some degree, the aggressive deterrence strategies and other offensive tactics
have had a corrosive effect upon shared cyber norms and their codification into
international and domestic laws~\citep{TikkRingas15}. In intentional power
politics small states have usually the most to lose, and thus it is not
surprising that cyber norms have been pushed forward particularly by small
states. Although global generalizations are problematic at best~\citep{Solar20},
together with non-governmental organizations and other actors, small states have
generally acted as so-called ``norm entrepreneurs'' by trying to influence the
international cyber security policy discourse through their normative
power~\citep{Adamson19}. This ``norm persuasion'' strategy is analogous to that
often used by the European Union (EU), with varying degrees of
success~\citep{Chan20}. With regard to normative power and small states in
general, Waltz is surprisingly silent. Basically he merely asserts vaguely that
only ``by merging and losing their political identities can middle states become
superpowers'' \citep[p.~182]{Waltz79}. While the reference to the late 1970s
Europe is clear but implicit, the roles played by small states provide a clear
hypothesis for a cross-country analysis. Given the traditional alliance options
for small states~\citep{Amstrup76}, it can be expected~that:
\begin{description}
\item[\normalfont{H}$_3$]{\textit{Small states prefer BMC for cyber security and international politics thereto.}}
\end{description}

The small but visible steps in the normative ordering principles have been
accompanied by efforts at the international political arenas, including the
United Nations (UN) in particular. The specific UN playgrounds have been the
International Telecommunication Union (ITU) and the so-called Group of
Governmental Experts (UNGGE) on Developments in the Field of Information and
Telecommunications in the Context of International Security. While the
UNGGE has been able to agree that international law applies to the cyber space,
otherwise disagreements have been widespread. Not only have there been conflicts
about definitions and scope, but the negotiations have also been affected by
hidden agendas and a disconnect from those who conduct offensive cyber
operations~\citep{Maurer20, Urgessa20}. Economic and trade issues have stirred
the pot further~\citep{Pomfret20}. Partially due to these deadlocks, other
governance forums have gained more traction, including those orchestrated by
companies \citep{vanHorenbeeck18}. All in all, it can be summarized that there
is at least a normative push toward a (horizontal) hierarchical order in the
international cyber space. At the same time, cyber capabilities continue to grow
and vary between states, and cyber power is actively maintained, exercised, and
demonstrated. It can be left to a reader to assess whether this state of
international cyber security affairs is an anarchy.

\section*{Data and Methods}

The empirical material is based on the so-called ``cyber power index'' (CPI)
dataset released by associates at the Harvard
University~\citep{CyberPower20Data}. Although \citet{ITU19a} has released a
comparable index, which has also been used in previous research \citep{Hansel17,
  Makridis19}, the CPI dataset has a couple of benefits speaking for itself. As
was already remarked, reliability and validity often remain issues for
macro-political indices, and the CPI dataset is not an exception in this
regard. However, first, the dataset is accompanied with a fairly detailed
codebook containing mostly complete references to the primary sources from which
the indices have been constructed. These sources seem reasonable. Unlike ITU,
second, the CPI dataset provides the individual variables from which the
composite ``cyber power index'' has been constructed. Originally, the
theoretical idea behind the dataset was to correlate the ``intent'' to use cyber
power against the capability to exercise it~\citep{CyberPower20Paper}. But as
the preceding discussion has tried to explicitly argue and implicitly persuade,
such a crude analysis does not really do honor to the potential offered by the
CPI dataset.

Thus, two separate dependent variables are used to proxy BMC. The first is the
amount and quality of bilateral and multilateral cooperation agreements in the
cyber space. These include both formal agreements, including memberships in
regional and global organizations, and informal arrangements, such as joint
declarations and cooperation frameworks. The scale of this dependent variable is
continuous, higher values indicating more agreements and more formality. The
second dependent variable proxies a country's endorsement of cyber norms. It is
coded from eleven criteria that mostly address a country's participation in
international organizations~\citep{CyberPower20Data}. In addition to ITU's
committees and the UNGGE, these organizations include both ``talkshops'', such
as the Internet Governance Forum (IGF) and the Global Forum for Cyber
Expertise~(GFCE), and more technical cooperation avenues, such as the
International Organization for Standardization (ISO) and the International
Electrotechnical Commission (IEC). In addition, national cyber security
strategies and related commitments are accounted for in the coding. The range of
this variable is scaled to the unit interval.

\begin{table}[p!]
\centering
\caption{Independent Variables}
\label{tab: independent variables}
\begin{tabularx}{\linewidth}{llX}
\toprule
A. & \multicolumn{2}{l}{Cyber capability ($\textmd{H}_1$)} \\
\cmidrule{1-2}
~~& CPI variables & Description \\
\cmidrule{2-3}
& \texttt{militarystrategy} & A national cyber strategy detailing defensive and (or) offensive military capabilities in the cyber space, ranked by consistency of the strategy. \\
\cmidrule{3-3}
& \texttt{cybermilpeople} & Number of staff in military cyber forces. \\
\cmidrule{3-3}

& \texttt{cybercommand} & The presence of a national, centralized cyber command, ranked by years since establishment. \\
\cmidrule{2-3}
B. & \multicolumn{2}{l}{Cyber power ($\textmd{H}_2$)} \\
\cmidrule{1-2}
~~& CPI variables & Description \\
\cmidrule{2-3}
& \texttt{stateattack} & Number of publicly attributed, notable, and sophisticated state-sponsored cyber attacks. \\
\cmidrule{3-3}
& \texttt{attacksurveillance} & Defined analogously to the \texttt{stateattack} variable, but with a surveillance objective. \\
\cmidrule{3-3}
& \texttt{attackcontrol} & Like \texttt{stateattack}, but with a control objective. \\
\cmidrule{3-3}
& \texttt{attackintelligence} & Ibid., but with an intelligence objective. \\
\cmidrule{3-3}
& \texttt{attackcommercial} & Ibid., but with a commercial objective. \\
\cmidrule{3-3}
& \texttt{attackoffense} & Ibid., but with an offensive objective. \\
\cmidrule{2-3}
& \texttt{intentoffense} & A variable with a $[0,1]$ range; based on seven questions asking whether a country's cyber military planning, strategy documents, etc.~acknowledge the capability for destructive cyber operations, including the destroying or disabling of adversaries' infrastructures and capabilities. \\
\cmidrule{2-3}
C. & \multicolumn{2}{l}{Small states ($\textmd{H}_3$)} \\
\cmidrule{1-2}
~~& CPI variables & Description \\
\cmidrule{2-3}
& -- & A dummy variable scoring one for Estonia, Lithuania, Singapore, Sweden, and Switzerland. \\
\bottomrule
\end{tabularx}
\end{table}

The CPI dataset contains information about $29$ countries. By implication, the
empirical analysis pursued falls to the so-called ``small-N'' category of
comparative research; there are only a few observations but a large number of
variables. The usual problems follow. Regarding these problems, particularly
problematic is the operationalization of cyber capabilities used to postulate
$\textmd{H}_1$. As was already discussed, these capabilities originate from the
capabilities of ``total societies''~\citep{Verba67}, from education to
technology and economy. How to quantify something like science? Although there
are no right answers to such a question (and many would argue that the question
is absurd to begin with), the CPI dataset contains various generic indices, such
as the prevalence of e-commerce, patent applications, and Internet
adoption. Given the importance of public-private partnerships for cyber
security~\citep{Ruohonen20EJSR}, indices are available also for the number of
cyber security and surveillance companies. However, the small-N constraints
force a more limited but sharper focus. Thus, the capabilities are explicitly
restricted to the military cyber capabilities listed in Table~\ref{tab:
  independent variables}. Given the wording used for~$\textmd{H}_2$, the
concept of cyber power is likewise framed to state-sponsored cyber attacks and to
the demonstration of such power through publicly available military
strategies. Finally, there is again no correct way to define small
states~($\textmd{H}_3$). Nevertheless, based on a subjective evaluation, five
countries are classified as small states; these include also countries whose
national security strategies have long relied on neutrality and the avoidance of
formal military alliances.

As for computation and methods, basic statistical techniques suffice to examine
the three hypotheses postulated. The principal components analysis (PCA) is used
to construct the composite cyber capability and cyber power variables in
Table~\ref{tab: independent variables}. Product-moment correlations and the
ordinary least squares (OLS) regression are used to assess their linear relation
to the BMC variables~constructed.

\section*{Results}

\begin{figure}[th!b]
\centering
\includegraphics[width=12cm, height=12cm]{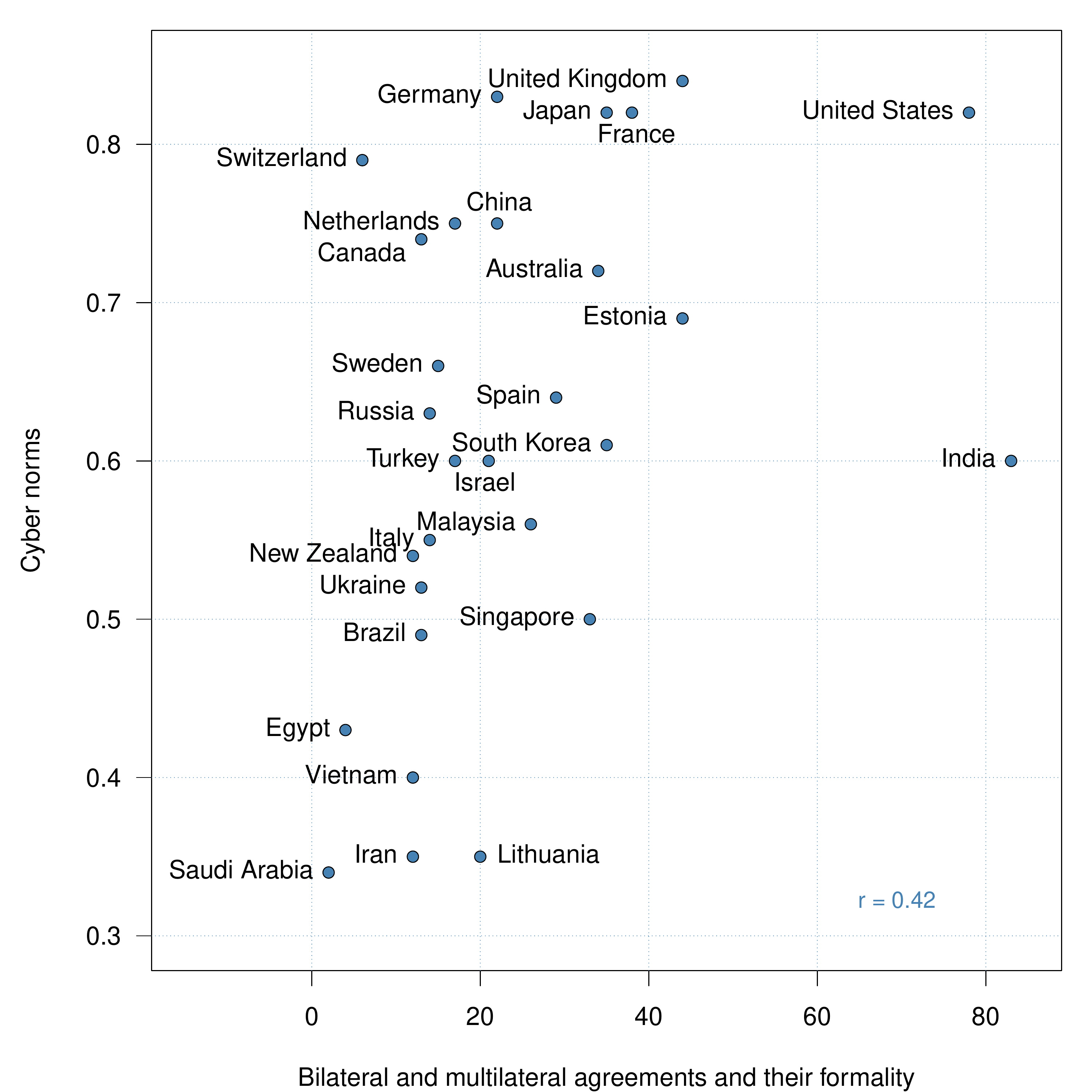}
\caption{The Two BMC Variables}
\label{fig: bmc}
\end{figure}

The variance between the $29$ countries is large with respect to their
commitments to BMC. As can be seen from Fig.~\ref{fig: bmc}, India and the
United States stand out as outliers in terms of bilateral and multilateral
agreements and their formality. Regarding the latter country, the result is
hardly unexpected~\citep{Solar20}. The United States ranks high also in terms of
cyber norms together with Japan and the three largest European countries. There
is a group of five countries that appear to have a lesser interest in the
norm-building activities. Among them is Lithuania. Even though Estonia ranks
fairly high in both BMC variables, a negative answer to $\textmd{H}_3$ therefore
seems plausible even without statistical computing. To some extent, these
cross-country observations support the arguments that many states still struggle
to integrate their cyber security strategies into their national (security)
strategies~\citep{Cavelty19a}. These struggles are not merely about rational
national strategies: cyber security politics are strongly present in many
countries; there are many ``swing states'' that have not decided on their vision
for the future of the cyber space and its security~\citep{Eldem20}. In any case,
the variance of both BMC variables is large enough for $\textmd{H}_1$ and
$\textmd{H}_2$ to be plausible.

A prerequisite for PCA and other related dimension reduction techniques is that
the target variables are correlated with each other~\citep{Seber84}. This
assumption provides a good way to start the empirical analysis. Thus, according
to \citeauthor{Bartlett51}'s \citeyearpar{Bartlett51} classical $\chi^2$-based
test the null hypothesis that a sample correlation matrix equals an identity
matrix is rejected for the six cyber power variables ($p < 0.001$) but not for
the three cyber capability variables ($p \simeq 0.094$). This observation hints
that the composite cyber capability variable might have problems with its
internal consistency. However, according to the PCA results, the first principal
component accounts for about 78\% of the total variance among the three cyber
capability variables. This is a reasonable share---and it is actually higher
than for the six cyber power variables (71\%). Therefore, the scores from the
first principal components can be reasonably used to examine Hypotheses~$\textmd{H}_1$ and
$\textmd{H}_2$.

\begin{table}[th!b]
\centering
\caption{Correlations (Pearson; $* \mapsto p < 0.05$, two-sided)}
\label{tab: cor}
\begin{tabular}{llrrrr}
\toprule
& & 1. & 2. & 3. & 4. \\
\cmidrule{3-6}
1. & Agreements~\\
2. & Norms & $0.42^*$ \\
3. & Capability & $-0.32^{\phantom{*}}$ & $-0.48^*$ \\
4. & Power & $-0.08^{\phantom{*}}$ & $0.12^{\phantom{*}}$ & $-0.32^{\phantom{*}}$ \\
5. & Small states & $0.04^{\phantom{*}}$  & $-0.06^{\phantom{*}}$  & $0.13^{\phantom{*}}$ & $-0.09^{\phantom{*}}$ \\
\bottomrule
\end{tabular}
\end{table}

\begin{table}[th!b]
\centering
\caption{Regression Results (OLS)}
\label{tab: reg 1}
\begin{tabular}{lrrcrr}
\toprule
& \multicolumn{2}{c}{Agreements ($R^2 = 0.139$)} && \multicolumn{2}{c}{Norms ($R^2 = 0.229$)} \\
\cmidrule{2-3}
\cmidrule{5-6}
Variable & Coef. & $p$-value && Coef. & $p$-value \\
\hline
Capability & $-0.003$ & $0.062$ && $< 0.001$ & $0.015$ \\
Power & $-0.444$ & $0.321$ && $-0.001$ & $0.848$ \\
Small states & $-0.222$ & $0.981$ && $-0.001$ & $0.991$ \\
\bottomrule
\end{tabular}
\end{table}

\begin{table}[th!b]
\centering
\caption{Auxiliary Regression Results (OLS)}
\label{tab: reg 2}
\begin{tabular}{lrrcrr}
\toprule
& \multicolumn{2}{c}{Agreements ($R^2 = 0.243$)} && \multicolumn{2}{c}{Norms ($R^2 = 0.582$)} \\
\cmidrule{2-3}
\cmidrule{5-6}
CPI variable & Coef. & $p$-value && Coef. & $p$-value \\
\hline
\texttt{scorecapabilities} & $0.989$ & $0.028$ && $0.009$ & $0.003$ \\
\texttt{scoreintent} & $-12.314$ & $0.653$ && $0.186$ & $0.263$ \\
\bottomrule
\end{tabular}
\end{table}

The correlations between the principal component scores and the two BMC
variables are shown in Table~\ref{tab: cor}. Although the coefficients are small
in magnitude for the cyber power composite, the cyber capability PCA-variable
correlates negatively with both the BMC agreements and cyber norms. For the
latter, the correlation coefficient is also statistically significant at the
conventional threshold. These observations are in accordance with the
prior expectation stated in the form of $\textmd{H}_1$.

However, the OLS regression results in Table~\ref{tab: reg 1} bring a dose of
skepticism. (The intercept is included in all regression models but not shown
for brevity.) Although the regression coefficient for the capability variable is
again statistically significant with respect to cyber norms, its magnitude is
extremely small. No other coefficient attains statistical significance. The
coefficients of determination are fairly decent for a three-variable model, but
still not particularly remarkable.

As there are more reasons to expect potential operationalization problems than
to assume that the statistical computation is faulty, four brief checks are in
order about the variables used. The first check is about the European Union,
which has had a strong influence upon the domestic cyber security laws in its
member states~\citep{Carvalho20, TikkRingas15}, and, to a lesser extent, upon
international cyber norms. However, replacing the dummy variable of the small
states with a dummy variable for the EU member states does not change the
results notably. The second check is similar but with a dummy variable for the
members of the North Atlantic Treaty Organization (NATO). Again, this dummy
variable is not statistically significant for either BMC variable. The third
check is about three large states (or, rather, active and powerful states in the
cyber space; China, Russia, and the United States), but, yet again, statistical
significance is not present. The fourth and final check is about the original
operationalization used in the CPI dataset. Thus, Table~\ref{tab: reg 2} shows a
simplified regression using only two independent variables: the ``intent'' to
use cyber power and another construction for the capability to exercise it
\citep{CyberPower20Paper}. The coefficient for the CPI's capability score
(\texttt{scorecapabilities}) is again statistically significant with respect to
cyber norms, but, again, the magnitude of it is small. Although the cyber norm
estimates generally yield a decent model ($R^2 \simeq 0.58$), the statistical
evidence does not seem convincing enough to outright accept
Hypothesis~$\textmd{H}_1$. That said, outright rejection seems appropriate for
Hypotheses~$\textmd{H}_2$ and $\textmd{H}_3$. These conclusions provide
interesting material for a brief speculation about the meaning behind the
results.

\section*{Conclusion}

The conclusion is easy to summarize: defensive and offensive cyber power and
capabilities for such power seem to neither increase nor decrease the
willingness and incentives of nation-states to participate in bilateral and
multilateral cooperation efforts on cyber security. The three hypotheses that
were contemplated can all be rejected. The statistical evidence falls somewhere
between ``no evidence of effect'' and effect ``too small to be worthwhile
pursuing'' (cf.~\citealt{daSilva15}) further. Something else drives the
incentives. But before any speculation should come a few words about the
so-called ``negative results''.  In general, these are ``non-results'' that fail
to provide a favorable outcome to prior research questions. Such results are
increasingly seen as valuable as they provide evidence on what is not yet known;
what remains unknown; what does not work. Such results are necessary for the
progress and integrity of science~\citep{daSilva15, Lehrer07}. Against this
backdrop, any result depends on the sincerity of researchers, and any negative
result can be turned into a ``positive result'' by insincere researchers merely
by rewriting the prior theories and research questions to match the results.

But how persuasive and realistic was the brief theorization along the
\citeauthor{Waltz79}'s \citeyearpar{Waltz79} classical take on structural
realism in international relations? Maybe the prior speculations and hypotheses
should have been rewritten after all? Without attempting to participate in the
extensive and everlasting debate about the Waltz's theory in general~(see
\citealt{Blachford20, Lechner17}, among many others), it is fairly easy to
attack his theory when it comes to cyber security. For years and years on, cyber
security has been said to differ from ``conventional'' security, and there are
no reasons to question this anthem, at least not yet. Thus, as an example, the
CPI dataset only accounts for things that are controllable by
states~\citep{CyberPower20Paper}, despite the fact that criticism has long been
levered against such state-centric approaches due to their omission of the
private sector, civil society, and many related aspects~~\citep{Mueller19,
  Ruohonen20EJSR}. Furthermore, can there even be a Waltzian balance of power in
cyber space?  To be sure, there are initiatives and programs to govern offensive
cyber technologies possessed by states~\citep{Herpig18}, but does counting these
really compare to counting nuclear warheads? But instead of attacking others,
perhaps a better option is to accept the negative empirical result presented and
contemplate the reasons behind it.

As was remarked, it is impossible to argue that there would not be incentives
for BMC, whether in terms of formal cooperation arrangements or cyber
norms. Surely all those committees and talkshops were not established for
nothing? To better understand the question, it is helpful to distinguish cyber
\textit{security politics} from cyber security \textit{politics}, the latter
referring to the politics engaging with cyber security
broadly~\citep{Cavelty19a}, or, rather, to the art of making cyber security
politics. From this viewpoint, the failure to establish solid norms and laws in
the cyber space would be a failure in politics. And, indeed, for many
practitioners, observers, and scholars alike~\citep{Register20a, Maurer20},
failures in diplomacy are the first thing coming to \textit{mind} from the
seemingly never-ending sequence of talkshops. To add a little bit of cynicism,
or realism, another thing coming to \textit{mind} is a facade. The CPI dataset
is enough to show that sophisticated state-sponsored cyber attacks continue
unabated. At the same time, cyber norms have been partially hijacked to advance
aggressive norms such as deterrence. While it is difficult to evaluate the
potential for an escalation, this kind of a thinking crosses the Rubicon. With
the helpful rereading of Waltz and others by others~\citep{Guzzini04}, it seems
fair to ask whether state actors in the cyber space are maximizing power or
whether they are maximizing security, and what are the implications if increases
in power do not translate into increases in~security?

Finally, it could well be that also the fundamental concepts were defined
incorrectly What if maximizing power is the ultimate goal and power is about the
``control over the minds and actions of others''? As for those whose memory
still serves them, there were realists long before Waltz, and there were also
those who equated political power to psychological power. To be sure, destroying
or disabling a digital infrastructure is a grievous act, but does it compare to
influencing \textit{total societies} via psychological means? Who was he who a
long time ago stated that political power is about the ``control over the
\textit{minds} and actions of others''?

\bibliographystyle{apalike}

\end{document}